\documentclass{aa}

\usepackage{graphicx}
\usepackage{natbib}
\begin{document}

\title{On the minimum and maximum mass of neutron stars and the delayed 
collapse}

\author{Klaus Strobel \and Manfred K. Weigel}

\offprints{K. Strobel, \email{Klaus.Strobel@physik.uni-muenchen.de}}

\institute{Sektion Physik, Ludwig-Maximilians Universit{\"a}t 
  M{\"u}nchen, Am Coulombwall 1, D-85748 Garching, Germany}

\date{Received July 27, 2000 / Accepted .......}

\titlerunning{On the minimum and maximum mass of neutron stars and the delayed 
collapse}
\authorrunning{K. Strobel \& M.K. Weigel}

\abstract{
The minimum and maximum mass of protoneutron stars and neutron stars
are investigated.  The hot dense matter is 
described by relativistic (including hyperons) and non-relativistic 
equations of state.
We show that the minimum mass ($\sim$\,0.88\,-\,1.28\,$M_{\sun}$) 
of a neutron star is determined by the earliest stage of 
its evolution and is nearly unaffected by the presence of hyperons.
The maximum mass of a neutron star is limited by the protoneutron star 
or hot neutron star stage.
Further we find that the delayed collapse of a neutron star into a black hole 
during deleptonization is not only possible for equations of state 
with softening components, as for instance, hyperons, meson 
condensates etc., but also for neutron stars with a pure 
nucleonic-leptonic equation of state.
}

\maketitle
%
\section{Introduction} \label{sec1}

It is expected that a neutron star (NS) is born as a result of the 
gravitational collapse of the iron core of a massive ($M\,>\,8 M_{\sun}$)
evolved star \cite[e.g.][]{Bet90}.
Shortly after core bounce (some 10~ms) a hot, lepton rich NS, called 
protoneutron star (PNS), is formed. This PNS evolves in some minutes 
into a cold, lepton pure NS due to the loss of neutrinos 
\cite[e.g.][]{BL86, KJ95, PRPLM99}. 
During these first minutes the PNS can collapse delayed to a black hole (BH) 
if its initial mass is high enough, either due to the loss of neutrinos 
or due to post bounce accretion.

The aim of this work is to study the minimum and the maximum mass 
of NSs and the possibility of BH formation during the deleptonization 
period. Recently \cite{Gon97, Gon98}, \cite{GHZ98} and \cite{Str99b} 
have calculated limits on the minimum mass of NSs, using equations 
of state (EOSs) including nucleons (n, p) and leptons. 
In this paper refinements to these approaches are given by calculating 
PNSs for a large sample of EOSs using a non-relativistic Thomas-Fermi 
model \cite[]{MS96, Str99a, Str99b} and a relativistic Hartree model 
\cite[]{SW86, Schaefer97} including hyperons.
Limits on the maximum mass of NSs were studied by numerous authors 
\cite[e.g.][]{Tak95, Bom96, ELP96, Pra97, Gon98} using different kinds 
of EOSs. All these investigations start with models of PNSs about 1\,-\,3~s 
after core bounce (see Sect.~\ref{sec2} for more details). 
In this paper the earliest stage of a PNS is taken 
additionally into account, 
which is reached about 100~ms after core bounce, to calculate the 
maximum mass of a NS (see Sect.~\ref{sec2} for more details).

\section{Inside a protoneutron star}
\label{sec2}

As pointed out in the Introduction a special topic of this investigation 
is the more detailed incorporation of the earliest stage of the 
PNS, about 0.1\,-\,1~s after core bounce 
\cite[]{BL86, BHF95, KJM96, PRPLM99}. 
This early type PNS is characterized by a hot
shocked envelope with an entropy per baryon of $s\,\sim$\,4\,-\,6 
(in units of the Boltzmann constant, $k_\mathrm{B}$) for baryon number 
densities $n\,<\,n_\mathrm{env}$, an unshocked core with $s\,\sim$\,1\,-\,2 for
densities $n\,>\,n_\mathrm{core}$, and a transition region between these
densities \cite[]{BHF95, KJM96}. 
For the description of this early stage of the PNS we use, in 
accordance with the investigations of \cite{BL88}, \cite{BHF95} 
and \cite{KJM96}, baryon densities in the range of 
0.002\,-\,0.02\,fm$^{-3}$ for $n_\mathrm{env}$ and 0.1\,fm$^{-3}$ for 
$n_\mathrm{core}$. The entropy per baryon is chosen between 5\,-\,6 
in the envelope and between 1\,-\,2 in the core, respectively. 
The investigation is performed for the early type PNS models with 
trapped neutrinos using a constant lepton number ($Y_\mathrm{l}\,=\,0.4$) 
for densities above $n\,=\,6\,\times\,10^{-4}$\,fm$^{-3}$. 

We assume that post bounce accretion onto the protoneutron star 
stops within the first 500\,ms after core bounce. The ammount 
of accreted matter, and the time at which the accretion through the 
shock front stops are still open questions. 
The amount of accreted matter after core bounce ranges from 
$0.5\,M_{\sun}$ during the first 10~ms or so to 
0.001 to $0.15\,M_{\sun}$ during the following 100~ms or so, 
but in most calculations significant accretion stops about 500~ms 
after core bounce 
\cite[for a further discussion of this topic see e.g.][]{BL88, Che89, 
HBC92, BHF95, JM96, Mez98b, FH99}.

About 1\,-\,3~s after core bounce the PNS has a nearly constant entropy per 
baryon ($s\,\,\sim\,2$) and an approximately constant lepton number 
($Y_\mathrm{l}\,\sim\,0.4$) for densities
$n\,>\,6\,\times\,10^{-4}$\,fm$^{-3}$.  The next stage in the lifetime of a NS
is reached after about 10\,-\,30~s,  where the hot NS ($s\,\sim\,2$) is now
transparent to neutrinos.  During the following minutes this hot NS evolves
into a cold NS.  For a more detailed description of the evolution of PNSs see,
for  instance, \cite{Pra97} and \cite{Str99b}.

The EOSs used in this paper for the description of PNSs are: 
(i) A non-relativistic Thomas-Fermi model (TF) for finite temperatures 
developed by \cite{Str99b, Str99a}. 
(ii) A relativistic Hartree model for finite temperatures 
developed by \cite{Schaefer97}.
The Hartree model includes the following hyperons and nucleonic 
isobars: $\Lambda, \Sigma^-, \Sigma^0, \Sigma^+, \Xi^-, \Xi^0, \Delta^-, 
\Delta^0, \Delta^+, \Delta^{++}$ (the Delta resonances are not 
included in the NL1 model).

\begin{figure}
  \resizebox{\hsize}{!}{\rotatebox{-90}{\includegraphics{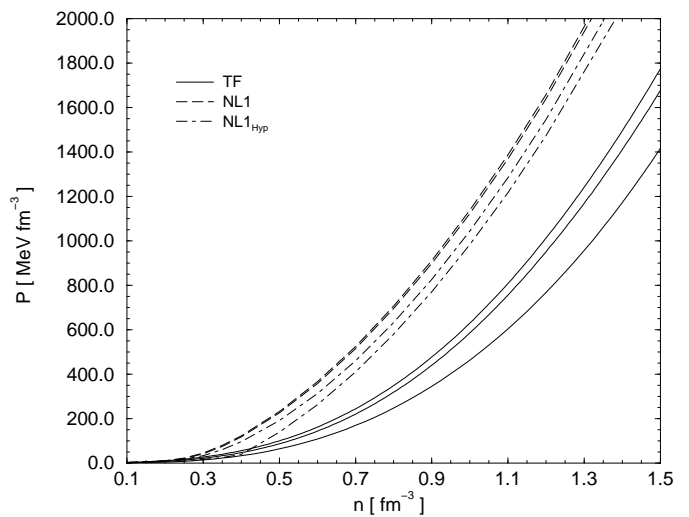}}}
  \caption[]{Pressure versus baryon number density in the high density region 
             for different EOSs. The solid curves 
             correspond (from the bottom to the top) to the cold EOS, the
             $s\,=\,1, Y_\mathrm{l}\,=\,0.4$ EOS and the $s\,=\,2, 
             Y_\mathrm{l}\,=\,0.4$ EOS of the TF model.  
             The long dashed curves correspond to the 
             cold EOS (lower) and the $s = 1, 
             Y_\mathrm{l}\,=\,0.4$ EOS (upper) of the NL1 model without 
             hyperons (dot-dashed lines, hyperons included).}
  \label{Pn}
\end{figure}

\begin{figure}
  \resizebox{\hsize}{!}{\rotatebox{-90}{\includegraphics{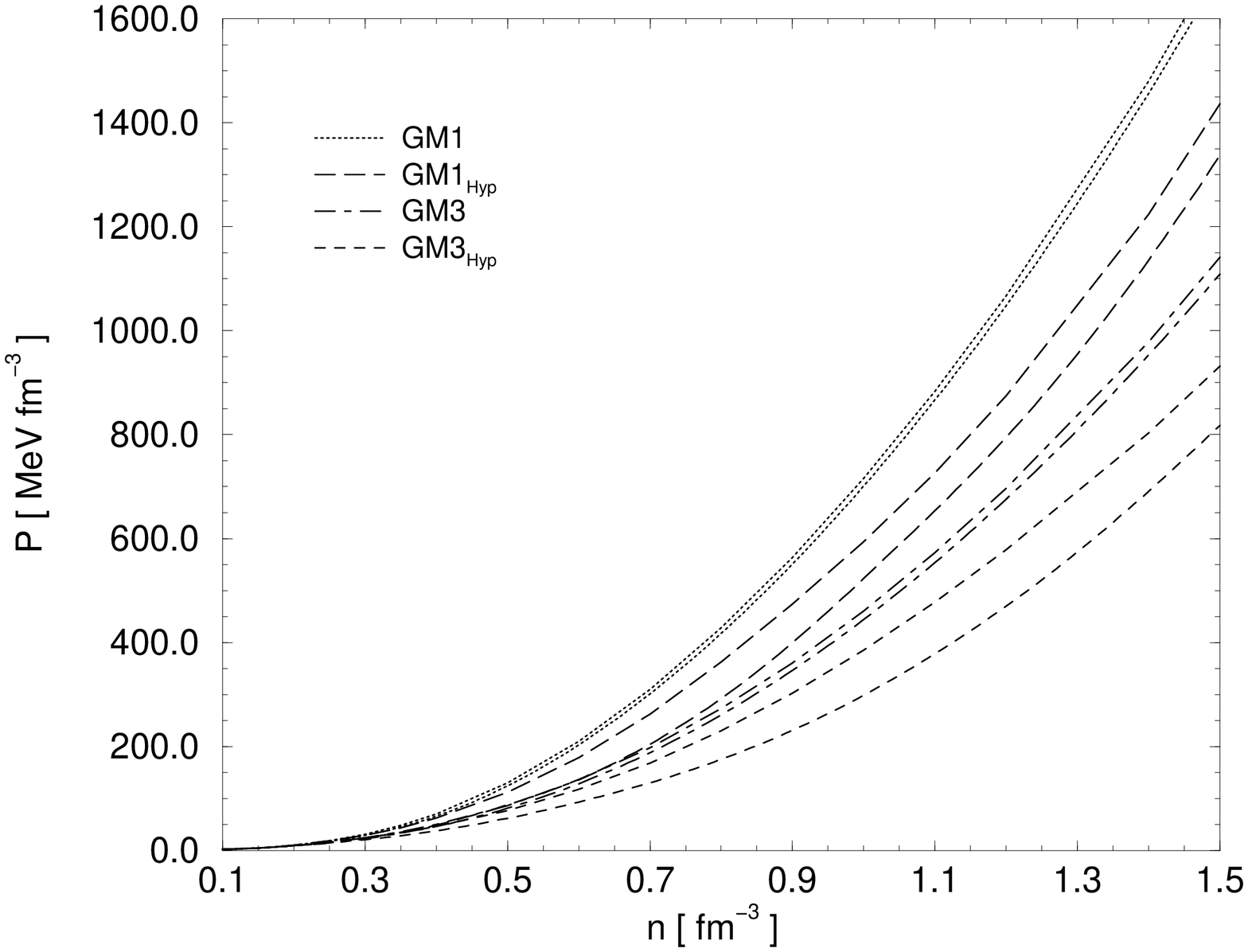}}}
  \caption[]{Pressure versus baryon number density in the high density region 
             for different EOSs. The dotted curves 
             correspond to the cold EOS (lower) and the 
             $s\,=\,1, Y_\mathrm{l}\,=\,0.4$ EOS (upper) of the GM1 
             model without hyperons (long dashed lines, hyperons included). 
             The dot-dashed curves correspond to the 
             cold EOS (lower) and the $s = 1, 
             Y_\mathrm{l}\,=\,0.4$ EOS (upper) of the GM3 model without 
             hyperons (short dashed lines, hyperons included).}
  \label{Pn1}
\end{figure}

The main properties of cold symmetric nuclear matter 
of the different models are listed in 
Table~\ref{nucl} \cite[see][]{Str99b, GM91, Rei89}.
The coupling constants for the Hartree 
models (GM1, GM3, NL1) are listed in Table~\ref{const}.
\begin{table}
\centering
  \caption[]{Properties of cold symmetric nuclear matter. 
The entries are: energy per baryon, $u$; saturation density, 
$n_0$; incompressibility, $K_\infty$; symmetry energy, $J$; effective 
nucleon mass, $m^*/m$.}
  \label{nucl}
  \begin{tabular}{ l c c c c c  }
  \hline
 & & & & & \\
    & $u$ & $n_0$ & $K_\infty$ & $J$ & $m^*/m$ \\
    & [MeV] & [fm$^{-3}$] & [MeV] & [MeV] &  \\

 & & & & & \\
  \hline
 & & & & & \\
   TF    & -16.24 & 0.161 & 234 & 32.7 &  0.87 \\
   GM1   & -16.30 & 0.153 & 300 & 32.5 &  0.70 \\
   GM3   & -16.30 & 0.153 & 240 & 32.5 &  0.78 \\
   NL1   & -16.42 & 0.152 & 212 & 43.5 &  0.57 \\
   
 & & & & & \\
  \hline
  \end{tabular} \\
\end{table}
\begin{table}
\centering
  \caption[]{Coupling constants for the relativistic Hartree models. 
The coupling constants are given for the following meson masses: 
$m_{\sigma}\,=\,550$~MeV, $m_{\omega}\,=\,783$~MeV and 
$m_{\rho}\,=\,770$~MeV (the coupling constants are converted according 
to these meson masses).}
  \label{const}
  \begin{tabular}{ l c c c c c  }
  \hline
 & & & & & \\
    & $g^2_\sigma/4\pi$ & $g^2_\omega/4\pi$ & $g^2_\rho/4\pi$ & 
    $10^3\bar b$ & $10^3\bar c$ \\
 & & & & & \\
  \hline
 & & & & & \\
   GM1   & 7.288 & 8.959 & 5.346 & 2.947 &  -1.070 \\
   GM3   & 6.139 & 6.041 & 5.807 & 8.659 &  -2.421 \\
   NL1   & 10.2099 & 13.6108 & 8.0260 & 2.4578 &  -3.4334 \\
 & & & & & \\
  \hline
  \end{tabular} \\
\end{table}
We take the nucleon-hyperon coupling strength equal to the 
nucleon-nucleon coupling strength for simplicity. 
This restriction will not change the results for the minimum mass 
of a NS, as we will show later.
The influence on the maximum mass will be strong, but for this 
point we refer to \cite{HWWS98}, where this question is studied 
for a large spectrum of different coupling constants and different 
nucleon-hyperon coupling strengths for cold NSs.
The EOSs for the different models with nucleonic-leptonic matter 
(TF, GM1, GM3 and NL1) and nuclear matter including hyperons 
(GM1$_\mathrm{Hyp}$, GM3$_\mathrm{Hyp}$ and NL1$_\mathrm{Hyp}$) are 
shown in Fig.~\ref{Pn} and Fig.~\ref{Pn1}. 
The EOSs for baryon number densities below $n\,=\,0.1$\,fm$^{-3}$ are taken 
from the investigation of \cite{Str99b}.

The NL1 coupling constants lead to the stiffest EOSs at high densities, 
since the high density behaviour is dominated by the strength of the 
$\omega$-coupling. However the incompressibility is the smallest. 
In this respect one has to keep in mind that the incompressibility 
determines the stiffness only near saturation and for higher densities 
the second and third derivatives of the incompressibility are important 
for the stiffness \cite[see for instance][]{RWW90}.

\section{The minimum mass of neutron stars}
\label{sec3}

The minimum and maximum mass of non-rotating NSs for the 
different EOSs and different 
choices of the transition region between the hot shocked envelope 
and the unshocked core of the PNS are given in Table~\ref{EOSs}.
It turned out that the minimum mass of a NS is determined by the 
earliest stage of the lifetime of a PNS. The values for the 
minimum gravitational (baryonic) mass, $M_\mathrm{G}$ ($M_\mathrm{B}$), 
lie in the range of 
0.878\,-\,1.284\,$M_{\sun}$ (0.949\,-\,1.338\,$M_{\sun}$) 
for the different models (see Table~\ref{EOSs}). 
\begin{table*}
\centering
  \caption[]{EOSs used in this paper and resulting minimum and maximum 
             masses.
             The entries are: time after core bounce, $t$;
             entropy per baryon in the 
             envelope, $s_\mathrm{env}$; 
             entropy per baryon in the 
             core, $s_\mathrm{core}$; 
             maximum baryon number density of the envelope correlated 
             with the 
             entropy per baryon in the envelope, 
             $n_\mathrm{env}(s_\mathrm{env}$); 
             minimum baryon number density of the core correlated 
             with the 
             entropy per baryon in the core, 
             $n_\mathrm{core}(s_\mathrm{core}$); 
             minimum baryonic mass of the PNS, $M_\mathrm{B}^\mathrm{min}$;
             resulting minimum gravitational mass of the cold NS,
             $M_\mathrm{G}^\mathrm{min}(T=0)$; 
             maximum baryonic mass of the PNS or NS, 
             $M_\mathrm{B}^\mathrm{max}$;
             maximum gravitational mass of the cold NS determined by the 
             PNS or hot NS, 
             $M_\mathrm{G}^\mathrm{max}(T=0)$;
             possibility of the formation of a black hole during 
             deleptonization, BH.}
  \label{EOSs}
  \begin{tabular}{ l c c c c c c c c c c  }
  \hline
 & & & & & & & & & & \\
  Model  & $t$ &
           $s_\mathrm{env}$ & $s_\mathrm{core}$ & 
           $n_\mathrm{env}(s_\mathrm{env})$ & 
           $n_\mathrm{core}(s_\mathrm{core})$ & 
           $M_\mathrm{B}^\mathrm{min}$ & 
           $M_\mathrm{G}^\mathrm{min}(T=0)$ & 
           $M_\mathrm{B}^\mathrm{max}$  & 
           $M_\mathrm{G}^\mathrm{max}(T=0)$ & BH \\
       & [s] &  &  & [fm$^{-3}$] & [fm$^{-3}$] & [$M_{\sun}$] & [$M_{\sun}$] &  
       [$M_{\sun}$] & [$M_{\sun}$] & \\
 & & & & & & & & & & \\
  \hline
   TF     & 0.1 - 1 & 6.0 & 2.0 & 0.002 & 0.1  & 1.338 & 1.284 & 2.366 
          & - & yes \\
          & 0.1 - 1 & 6.0 & 1.0 & 0.002 & 0.1  & 0.949 & 0.878 & 2.371 
          & - & yes \\
          & 0.1 - 1 & 6.0 & 1.0 & 0.005 & 0.1  & 1.117 & 1.021 & 2.372 
          & - & yes \\
          & 0.1 - 1 & 5.0 & 1.0 & 0.01  & 0.1  & 1.020 & 0.940 & 2.372 
          & - & yes \\
          & 0.1 - 1 & 5.0 & 1.0 & 0.02  & 0.1  & 1.229 & 1.114 & 2.374 
          & - & yes \\
  \cline{2-11}
          & 1 - 3 & 2.0 & 2.0 & -  & -  & - & - & 2.364 
          & 1.969 & - \\
          & 10 - 30 & 2.0 & 2.0 & -  & -  & - & - & 2.449 
          & - & - \\
          & $\infty$ & 0.0 & 0.0 & -  & -  & - & - & 2.417
          & - & - \\
   \hline
   GM1    & 0.1 - 1 & 6.0 & 1.0 & 0.005 & 0.1 & 1.255 & 1.206 & 2.629 
          & - & yes \\
          & 0.1 - 1 & 5.0 & 1.0 & 0.005 & 0.1 & 1.019 & 0.988 & 2.628 
          & - & yes \\
          & 0.1 - 1 & 5.0 & 1.0 & 0.01  & 0.1 & 1.140 & 1.100 & 2.629 
          & - & yes \\
  \cline{2-11}
          & 1 - 3 & 2.0 & 2.0 & - & - & - & - & 2.613 
          & 2.251 & - \\
          & 10 - 30 & 2.0 & 2.0 & -  & - & - & - & 2.670 
          & - & - \\
          & $\infty$ & 0.0 & 0.0 & -  & - & - & - & 2.706
          & - & - \\
   \hline
   GM1$_\mathrm{Hyp}$    & 0.1 - 1 & 6.0 & 1.0 & 0.005 & 0.1 & 1.255 & 1.206 
                         & 2.462 & - & yes \\
                         & 0.1 - 1 & 5.0 & 1.0 & 0.005 & 0.1 & 1.019 & 0.987 
                         & 2.461 & - & yes \\
                         & 0.1 - 1 & 5.0 & 1.0 & 0.01  & 0.1 & 1.140 & 1.100 
                         & 2.462 & - & yes \\
  \cline{2-11}
                         & 1 - 3 & 2.0 & 2.0 & - & - & - & -
                         & 2.401 & - & yes \\
                         & 10 - 30 & 2.0 & 2.0 & - & - & - & - 
                         & 2.304 & 2.015 & - \\
                         & $\infty$ & 0.0 & 0.0 & - & - & - & - 
                         & 2.353 & - & - \\
   \hline
   GM3    & 0.1 - 1 & 6.0 & 1.0 & 0.005 & 0.1 & 1.255 & 1.205 & 2.204 
          & 1.939 & no \\
          & 0.1 - 1 & 5.0 & 1.0 & 0.005 & 0.1 & 1.039 & 1.005 & 2.203 
          & 1.938 & no \\
          & 0.1 - 1 & 5.0 & 1.0 & 0.01  & 0.1 & 1.145 & 1.104 & 2.204  
          & 1.939 & no \\
  \cline{2-11}
          & 1 - 3 & 2.0 & 2.0 & - & - & - & - & 2.212 
          & - & - \\
          & 10 - 30 & 2.0 & 2.0 & -  & - & - & - & 2.249 
          & - & - \\
          & $\infty$ & 0.0 & 0.0 & -  & - & - & - & 2.247 
          & - & - \\
   \hline
   GM3$_\mathrm{Hyp}$    & 0.1 - 1 & 6.0 & 1.0 & 0.005 & 0.1 & 1.255 & 1.205 
                         & 2.047 & - & yes \\
                         & 0.1 - 1 & 5.0 & 1.0 & 0.005 & 0.1 & 1.039 & 1.005 
                         & 2.046 & - & yes \\
                         & 0.1 - 1 & 5.0 & 1.0 & 0.01  & 0.1 & 1.145 & 1.104 
                         & 2.047 & - & yes \\
  \cline{2-11}
                         & 1 - 3 & 2.0 & 2.0 & - & - & - & -
                         & 2.015 & - & yes \\
                         & 10 - 30 & 2.0 & 2.0 & -  & - & - & - 
                         & 1.898 & 1.699 & - \\
                         & $\infty$ & 0.0 & 0.0 & -  & - & - & - 
                         & 1.911 & - & - \\
  \hline
  NL1    & 0.1 - 1 & 5.0 & 1.0 & 0.005 & 0.1 & 1.175 & 1.111 & 3.218 
          & - & yes \\
  \cline{2-11}
          & 1 - 3 & 2.0 & 2.0 & - & - & - & - & 3.162 
          & 2.663 & (yes) \\
          & 10 - 30 & 2.0 & 2.0 & -  & - & - & - & 3.202 
          & (2.689) & - \\
          & $\infty$ & 0.0 & 0.0 & -  & - & - & - & 3.297 
          & - & - \\
   \hline
   NL1$_\mathrm{Hyp}$    & 0.1 - 1 & 5.0 & 1.0 & 0.005 & 0.1 
                         & 1.169 & 1.104 & 3.017 & - & yes \\                  
      \cline{2-11}   
          & 1 - 3 & 2.0 & 2.0 & - & - & - & - & 2.942 & - & yes \\            
             & 10 - 30 & 2.0 & 2.0 & -  & - & - & - & 2.871 & 2.399 & - \\    
            & $\infty$ & 0.0 & 0.0 & -  & - & - & - & 2.962 & - & - \\ 
  \hline
  \end{tabular} \\
\end{table*}
They are larger by a factor of ten than the minimum mass of a cold NS 
\cite[e.g.][]{Str97} emerging from a calculation for the minimum mass 
of the cold NS based only on the solution of the Tolman-Oppenheimer-Volkoff 
equation with the corresponding EOS for cold 
NS matter. This shift due to the properties of early type PNSs was also 
recently found by \cite{GHZ98} and \cite{Str99b}. \cite{Gon97, Gon98} 
found an even smaller value for the minimum mass 
($M_\mathrm{G}\,\sim\,0.6\,M_{\sun}$). 
The reason for this is that they use models for PNSs about 1\,-\,3~s
after core bounce, where the minimum mass is lower, see also 
\cite{Str99b}. 
Our models with hyperons included do not change the values for the minimum 
mass significantly ($\Delta\,M_\mathrm{G}\,\leq\,0.007 M_{\sun}$). 
This result could be expected since the maximum baryon number density inside 
the core of such a NS is smaller than two times the nuclear matter density 
and hence the hyperon fraction is rather small. 

Let us now turn to the question, why there is a minimum mass 
in this early stage of the evolution.
A star becomes dynamically unstable if its mean adiabatic index, 
$\bar{\Gamma}$, lies below 4/3. 
In the case of a cold NS this is caused by the neutron drip 
at a density of $n\,\sim\,2.6 \times 10^{-4}$~fm$^{-3}$.
For the earliest stage of the PNS the drop below 
$\bar{\Gamma}\,=\,4/3$ is caused by the entropy drop between 
the hot shocked envelope and the unshocked core of the PNS.
For a complete description of the stability criteria of 
neutron stars see \cite{ST83}.

\section{The maximum mass of neutron stars and the 
possibility of black hole formation during deleptonization} \label{sec3a}

\begin{figure}
  \resizebox{\hsize}{!}{\rotatebox{-90}{\includegraphics{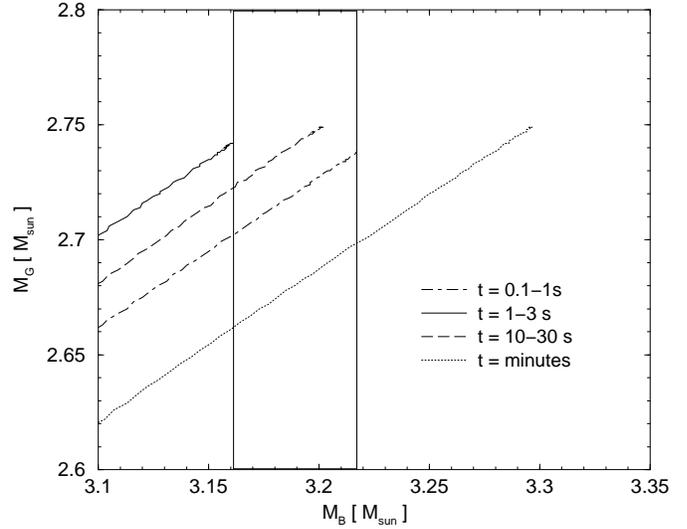}}}
  \caption[]{Maximum gravitatinal versus maximum baryonic mass for 
             different evolutionary stages of the NL1 EOSs. 
             The two lines denote the mass window for the delayed collapse.}
  \label{mbmg}
\end{figure}

Quite interesting is the result for the maximum gravitational mass for 
NSs, for which we obtain values in the range between 
1.699\,-\,2.663\,$M_{\sun}$ (see Table~\ref{EOSs}) for the different 
EOSs used in this paper, with the possibility 
of the formation of a BH during deleptonization for EOSs with a pure 
nucleonic-leptonic composition and EOSs which include hyperons. 
For instance, for the NL1 EOS, the maximum baryonic mass for the first 
milliseconds, $M_\mathrm{B}^\mathrm{max}\,=$\,3.218\,$M_{\sun}$, is 
larger than the allowed maximum baryonic mass of 3.162 $M_{\sun}$ after
1\,-\,3~s,  which implies the possibility of a BH formation 
(see Table~\ref{EOSs} and Fig.~\ref{mbmg}). 
This value also sets the boundary of 2.663\,$M_{\sun}$ for the maximum 
gravitational mass for the cold NS. 
The maximum baryonic mass of the first stage is even larger than 
maximum value 10\,-\,30~s after core bounce fore the NL1 EOS. 
This means that the delayed collapse is also possible for 
the totally deleptonized hot NS of the NL1 EOS. 
It should be mentioned in this context that the exact entropy profile 
of the hot shocked envelope of an early type PNS 
plays a minor role in determining 
the maximum baryonic mass, since a PNS with a constant entropy per 
baryon of $s\,=\,1$ throughout the whole PNS and a lepton number of 
$Y_\mathrm{l}\,=\,0.4$ leads to nearly the same maximum baryonic mass 
(e.g. for the NL1 model: $M_{\mathrm{B}}^{\mathrm{max}}\,=\,3.218\,M_{\sun}$, 
0.1\,-\,1.0~s after core bounce; 
$M_{\mathrm{B}}^{\mathrm{max}}\,=\,3.219\,M_{\sun}$, for an entropy per 
baryon of $s\,=\,1$ throughout the whole PNS and $Y_\mathrm{l}\,=\,0.4$). 
This means that the maximum mass depents nearly complete on the 
properties of the core of the PNS\footnote{Note: Other properties of an 
early type PNS (e.g. the radius) are strongly affected by the 
entropy profile.}.

\begin{figure}
  \resizebox{\hsize}{!}{\rotatebox{-90}{\includegraphics{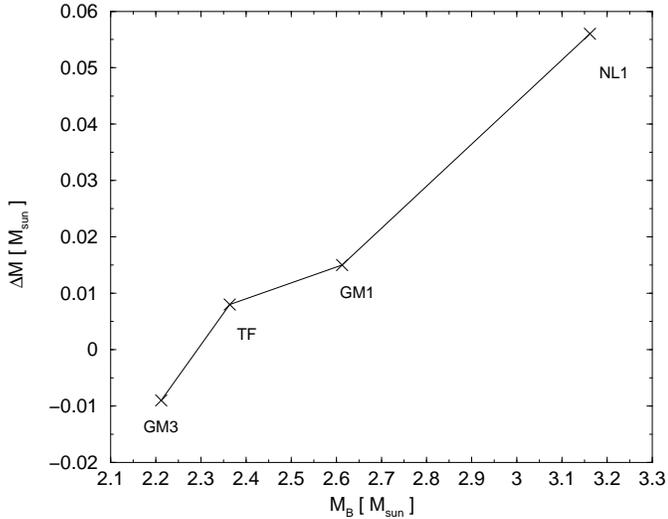}}}
  \caption[]{Baryonic mass difference between the early type PNS stage 
(0.1\,-\,1~s after core bounce) and the late type PNS stage 
(1\,-\,3~s after core bounce) versus resulting maximum baryonic 
mass for the different pure nucleonic-leptonic EOSs.}
  \label{deltam}
\end{figure}

Similar behaviour is found for the other EOSs with exception of the 
nucleonic-leptonic EOS GM3, where the allowed maximum mass after 
1\,-\,3~s exeeds the maximum masses of the first period. 
The reason why this possibility of a BH formation was dismissed so far 
for a pure nucleonic-leptonic composition \cite[e.g.][]{Tak95, Bom96, 
ELP96, Pra97, Gon98} was that these authors start their calculations 
about 1\,-\,3~s after core bounce, since the following stages 
allow only larger baryonic masses. 
One should mention in this context that for the nucleonic-leptonic 
EOSs the differences in the maximum masses in the first seconds are more than 
three times smaller\footnote{For that reason the effect was overlooked in 
\cite{Str99b}.} ($<\,0.056\,M_{\sun}$) than for the softer EOSs with 
hyperons (0.146\,-\,0.158 $M_{\sun}$). Due to this, 
for the EOSs including hyperons the possibility of a BH 
formation shows up even by starting the calculation  
1\,-\,3~s after core bounce (see Table~\ref{EOSs}). 
It turned out, that the mass window for the delayed collapse 
of the pure nucleonic-leptonic EOSs is an inceasing function of the maximum 
possible mass of the different models (see Fig.~\ref{deltam}).
The EOSs including hyperons did not show a similar behaviour, 
the mass window is nearly constant for all used EOSs.
The minimum value of the maximum baryonic mass for the models 
including hyperons is reached by the hot NS without trapped neutrinos 
(10\,-\,30~s after core bounce).
This, as already mentioned, can lead to a delayed collapse of the NS 
during deleptonization if the initial mass of the PNS is high enough 
\cite[e.g.][]{BJKST96, PRPLM99}. 

It is an open question how large the maximum mass 
of a NS really is. It is clear that an EOS should allow NS masses 
larger than $1.4\,M_{\sun}$ since measurements of pulsars in 
binary systems show up values around this mass \cite[e.g.][]{TC99}. 
If quasi periodic oscillations \cite[e.g.][]{vdK00}
are taken into account, the possible 
maximum mass of an EOS  has to be larger than $1.8\,-\,2.0\,M_{\sun}$
\cite[e.g.][]{Schaab99}. This whould lead to the conclusion that only 
the GM1 and the NL1 EOSs (with and without hyperons) are 
stiff enough to allow NSs with a larger mass.
\section{Discussion and conclusion} \label{sec4}

In this paper we presented a calculation of the minimum and maximum mass of 
a NS taking the early evolution, shortly after core bounce, into account.
We found that the minimum mass of a NS is limited by the earliest stage 
of its evolution ($\sim$\,0.1\,-\,1~s after core bounce). Therefore 
the minimum gravitational mass has values between 0.878 and 1.284 
$M_{\sun}$. These values lie in the same range as those found in recent works 
by \cite{GHZ98} and \cite{Str99b}. Additionally we have examined the 
influence of hyperons on the minimum mass. As expected it turned out that the 
minimum mass is nearly unaffected by the inclusion of hyperons. 
Another possibility of NS production is the accretion induced collapse
of a white dwarf \cite[]{CS76}, but the numerical simulation of 
\cite{WB92} showed that the resulting NS will have a mass of 
$\sim\,1.4\,M_{\sun}$ since the white dwarf collapses without significant
mass loss after reaching the maximum mass of a white dwarf. 
From the results reported it is obvious that NSs, which are born in 
Type-II supernovae or due to accretion induced collapse of a white dwarf, 
have a minimum mass about ten times larger than 
the possible minimum mass of a cold NS ($\sim\,0.1 M_{\sun}$).

With respect to the maximum mass of a NS calculated by solving the 
Tolman-Oppenheimer-Volkoff equation for NS matter the outcome is 
rather uncertain, since the EOS in the high density region rests 
strongly on theoretical extrapolations. 
The underlying model, for instance, different many-body 
approximations and/or inclusion of more massive baryons, quarks, 
condensates etc., can render the results for the maximum masses 
significantly \cite[e.g.][]{Pra97, HWWS98}.
However, as a result of our work it turned out, that the maximum mass of a NS 
is limited by the first $\sim$ 30 seconds of its evolution. 
The maximum baryonic mass of a NS, constructed with pure nucleonic-leptonic 
EOSs and EOSs including hyperons, is always reached during this early epoch 
of the evolution.

Finally we found that BH formation during deleptonization is also possible 
for NSs with pure nucleonic-leptonic EOSs if the initial mass is high 
enough. This property of NSs was 
expected till now to be possible only for EOSs including softening 
ingredients, like hyperons, meson condensates or a quark-hadron phase 
transition \cite[e.g.][]{BB94, Gle95}. 
There are numerous papers on this topic 
\cite[e.g.][]{Tak95, Bom96, ELP96, Pra97, Gon98}, but all these 
investigations started their calculations about 1\,-\,3~s after core 
bounce. 
In this paper we included the earliest stage and found that the maximum 
baryonic mass of the early type PNSs ($\sim$\,0.1\,-\,1~s after core
bounce) could be up to 0.056\,$M_{\sun}$ larger than that of 
the late type PNS ($\sim$\,1\,-\,3~s after core bounce). 
Due to this result it seems possible that 
a NS with a pure nucleonic-leptonic EOS could also undergo delayed 
collapse into a BH during the first seconds of its evolution. This question 
could only be finally decided in time evolution calculations like this 
including softening ingredients \cite[e.g.][]{BJKST96}, since 
we have used approximate values for the entropy per baryon and the lepton 
number at fixed times. 
Post bounce accretion \cite[e.g.][]{Che89, HBC92} could lead also to a 
delayed collapse, but it is uncertain till now how much matter falls 
through the shock front within the first second after core bounce, 
so that if a cut off of a neutrino signal after a 
few seconds is detected in the future, the two ways of BH formation  
could possibly be indistinguishable.
\begin{acknowledgements}
We want to thank A. Sch\"afer for providing us with his program.
We want to thank H.-Th. Janka and Ch. Schaab for many helpful discussions.  
One of us, K.~S., gratefully acknowledges the Bavarian State for financial
support.
\end{acknowledgements}

\vspace{\fill}
This article was processed by the author using Springer-Verlag (revised 
by EDP Sciences) \LaTeX~A\&A style file L-AA version 5.01.


\begin{thebibliography}{38}
\expandafter\ifx\csname natexlab\endcsname\relax\def\natexlab#1{#1}\fi

\bibitem[Baumgarte et~al.(1996)Baumgarte, Janka, Keil, Shapiro, \&
  Teukolsky]{BJKST96}
Baumgarte T.W., Janka H.T., Keil W., Shapiro S.L., Teukolsky S.A., 1996, ApJ
  468, 823

\bibitem[Bethe(1990)]{Bet90}
Bethe H.A., 1990, Rev. Mod. Phys. 62, 801

\bibitem[Bombaci(1996)]{Bom96}
Bombaci I., 1996, ApJ 305, 871

\bibitem[Brown \& Bethe(1994)]{BB94}
Brown G.E., Bethe H.A., 1994, ApJ 423, 659

\bibitem[Burrows \& Lattimer(1986)]{BL86}
Burrows A., Lattimer J.M., 1986, ApJ 307, 178

\bibitem[Burrows \& Lattimer(1988)]{BL88}
Burrows A., Lattimer J.M., 1988, Phys. Rep. 163, 51

\bibitem[Burrows et~al.(1995)Burrows, Hayes, \& Fryxell]{BHF95}
Burrows A., Hayes J., Fryxell B.A., 1995, ApJ 450, 830

\bibitem[Canal \& Schatzman(1976)]{CS76}
Canal R., Schatzman E., 1976, A\&A 46, 229

\bibitem[Chevalier(1989)]{Che89}
Chevalier R.A., 1989, ApJ 346, 847

\bibitem[Ellis et~al.(1996)Ellis, Lattimer, \& Prakash]{ELP96}
Ellis P.J., Lattimer J.M., Prakash M., 1996, Comments Nucl. Part. Phys. 22, 63

\bibitem[Fryer \& Heger(2000)]{FH99}
Fryer C.L., Heger A., 2000, ApJ 541, 1033

\bibitem[Glendenning(1995)]{Gle95}
Glendenning N.K., 1995, ApJ 448, 797

\bibitem[Glendenning \& Moszkowski(1991)]{GM91}
Glendenning N.K., Moszkowski S.A., 1991, Phys. Rev. Lett. 67, 2414

\bibitem[Gondek et~al.(1997)Gondek, Haensel, \& Zdunik]{Gon97}
Gondek D., Haensel P., Zdunik J.L., 1997, A\&A 325, 217

\bibitem[Gondek et~al.(1998)Gondek, Haensel, \& Zdunik]{Gon98}
Gondek D., Haensel P., Zdunik J.L., 1998, in Chan K.L., Cheng K.S., Singh H.P.
  (eds.), 1997 Pacific Rim Conference on Stellar Astrophysics, ASP Conference
  Series, Vol. 138, p. 131

\bibitem[Goussard et~al.(1998)Goussard, Haensel, \& Zdunik]{GHZ98}
Goussard J.O., Haensel P., Zdunik J.L., 1998, A\&A 330, 1005

\bibitem[Herant et~al.(1992)Herant, Benz, \& Colgate]{HBC92}
Herant M., Benz W., Colgate S., 1992, ApJ 395, 642

\bibitem[Huber et~al.(1998)Huber, Weber, Weigel, \& Schaab]{HWWS98}
Huber H., Weber F., Weigel M.K., Schaab C., 1998, Int. J. Mod. Phys. E 7, 301

\bibitem[Janka \& M{\"u}ller(1996)]{JM96}
Janka H.T., M{\"u}ller E., 1996, A\&A 306, 167

\bibitem[Keil \& Janka(1995)]{KJ95}
Keil W., Janka H.T., 1995, A\&A 296, 145

\bibitem[Keil et~al.(1996)Keil, Janka, \& M{\"u}ller]{KJM96}
Keil W., Janka H.T., M{\"u}ller E., 1996, ApJ 473, L111

\bibitem[Mezzacappa et~al.(1998)Mezzacappa, Calder, Bruenn, et~al.]{Mez98b}
Mezzacappa A., Calder A.C., Bruenn S.W., et~al., 1998, ApJ 495, 911

\bibitem[Myers \& {\'S}wi{\c a}tecki(1996)]{MS96}
Myers W.D., {\'S}wi{\c a}tecki W.J., 1996, Nucl. Phys. A 601, 141

\bibitem[Pons et~al.(1999)Pons, Reddy, Prakash, Lattimer, \& Miralles]{PRPLM99}
Pons J.A., Reddy S., Prakash M., Lattimer J.M., Miralles J.A., 1999, ApJ 513,
  780

\bibitem[Prakash et~al.(1997)Prakash, Bombaci, Prakash, et~al.]{Pra97}
Prakash M., Bombaci I., Prakash M., et~al., 1997, Phys. Rep. 280, 1

\bibitem[Ramsch{\"u}tz et~al.(1990)Ramsch{\"u}tz, Weber, \& Weigel]{RWW90}
Ramsch{\"u}tz J., Weber F., Weigel M.K., 1990, J. Phys. G 16, 987

\bibitem[Reinhard(1989)]{Rei89}
Reinhard P.G., 1989, Rep. Prog. Phys. 52, 439

\bibitem[Schaab \& Weigel(1999)]{Schaab99}
Schaab C., Weigel M.K., 1999, M.N.R.A.S. 308, 718

\bibitem[Sch{\"a}fer(1997)]{Schaefer97}
Sch{\"a}fer A., 1997, ``Eigenschaften der Materie hei{\ss}er
  Protoneutronensterne'', Master's thesis, Universit{\"a}t M{\"u}nchen,
  unpublished

\bibitem[Serot \& Walecka(1986)]{SW86}
Serot B.D., Walecka J.D., 1986, Adv. Nucl. Phys. 16, 1

\bibitem[Shapiro \& Teukolsky(1983)]{ST83}
Shapiro S.L., Teukolsky S.A., 1983, ``Black Holes, White Dwarfs and Neutron
  Stars'', John Wiley \& Sons, New York

\bibitem[Strobel et~al.(1997)Strobel, Weber, Schaab, \& Weigel]{Str97}
Strobel K., Weber F., Schaab C., Weigel M.K., 1997, Int. J. Mod. Phys. E 6, 669

\bibitem[Strobel et~al.(1999{\natexlab{a}})Strobel, Schaab, \& Weigel]{Str99b}
Strobel K., Schaab C., Weigel M.K., 1999{\natexlab{a}}, A\&A 350, 497

\bibitem[Strobel et~al.(1999{\natexlab{b}})Strobel, Weber, \& Weigel]{Str99a}
Strobel K., Weber F., Weigel M.K., 1999{\natexlab{b}}, Z. Naturforsch. 54a, 83

\bibitem[Takatsuka(1995)]{Tak95}
Takatsuka T., 1995, Nucl. Phys. A 588, 365

\bibitem[Thorsett \& Chakrabarty(1999)]{TC99}
Thorsett S.E., Chakrabarty D., 1999, ApJ 512, 288

\bibitem[{van der Klis}(2000)]{vdK00}
{van der Klis} M., 2000, ARA\&A 38, 717

\bibitem[Woosley \& Baron(1992)]{WB92}
Woosley S.E., Baron E., 1992, ApJ 391, 228

\end{thebibliography}
\end{document}